\documentclass{article}
\usepackage{graphicx} 
\usepackage{placeins}
\usepackage[utf8]{inputenc}
\usepackage{newunicodechar}
\usepackage{csquotes}
\usepackage{siunitx}
\usepackage{booktabs}
\usepackage[german,english]{babel}
\usepackage[backend=biber,bibstyle = nature, maxbibnames=99, isbn=false,sorting=none,bibencoding=auto,intitle=true,urldate=long,eprint=false]{biblatex}
\usepackage{hyperref}
\DeclareUnicodeCharacter{2009}{\,}
\DeclareUnicodeCharacter{03BF}{o}
\AtEveryBibitem{%
  \clearfield{note}%
}

\title{Ultra-broadband UV/VIS spectroscopy enabled by resonant dispersive wave emission of a frequency comb}
\author{Adrian Kirchner}
\date{July 2024}

\begin{document}

\begin{flushleft}
{\Large
\textbf\newline{Ultra-broadband UV/VIS spectroscopy enabled by resonant dispersive wave emission of a frequency comb}
}
\newline
\\
Adrian Kirchner\textsuperscript{1},
Alexander Eber\textsuperscript{1},
Lukas Fürst\textsuperscript{1},
Emily Hruska\textsuperscript{1},
Michael H. Frosz\textsuperscript{2},
Francesco Tani\textsuperscript{2,3},
Birgitta Bernhardt\textsuperscript{1,*}
\\
\bigskip
\bf{1} Institute for Experimental Physics, Graz University of Technology, Petersgasse 16, 8010 Graz, Austria
\\
\bf{2} Max Planck Institute for the Science of Light, Staudtstrasse 2, 91058 Erlangen, Germany
\\
\bf{3} Univ. Lille, CNRS, UMR 8523-PhLAM-Physique des Lasers Atomes et Molécules, F-59000 Lille, France\\
*  Correspondence to: \href{mailto:bernhardt@tugraz.at}{bernhardt@tugraz.at}
\bigskip

\end{flushleft}

\section*{Abstract}
\label{sec:abstract}

We introduce a novel ultra-broadband ultraviolet and visible frequency comb light source covering more than $240\,\si{THz}$ by resonant dispersive wave emission in a gas-filled hollow-core fiber waveguide. The light source allows tuning from $\approx 340\,\si{nm}$ to $465\,\si{nm}$ ($645\,\si{THz}$ to $\approx 885\,\si{THz}$) with conversion efficiencies of $1.5\,\%$. Ultra-broadband absorption spectroscopy is demonstrated by studying nitrogen dioxide, a molecular species of major atmospheric relevance strongly absorbing across the ultraviolet and visible spectral region. We show that the coherence of the 80 MHz ytterbium fiber-based frequency comb seeding the frequency up-conversion process is conserved, paving the way toward further ultra-broadband (dual) comb spectroscopy across the ultraviolet/visible range.

\section{Introduction}
\label{sec:intro}

The ultraviolet (UV) spectral region is of high interest in various disciplines from fundamental scientific questions \cite{ding_threshold_2019,kotsina_spectroscopic_2022,lee_few-femtosecond_2024,ghezzi_design_2023,furst_broadband_2024} to atmospheric sciences  \cite{wahner_near_1990,platt_measurements_1983} and food science \cite{alamprese_detection_2013,urickova_determination_2015,weeranantanaphan_identity_2010,wlodarska_non-destructive_2019}. For these studies, various light sources are used, depending on the requirements of the application. Arc lamps offer broad spectral coverage but no coherence and low brightness \cite{hecht_optics_2017}. Synchrotrons provide high brightness, tuneability, and coherence up to MHz repetition rates with ps time resolution but their operation is expensive and beam time access is limited  \cite{codling_applications_1973,attwood_soft_2000}. Alternatively, nonlinear up-conversion of laser radiation can coherently access the UV range with brightness and temporal resolution dependent on the driving laser. However, these sources often lack either tunability or bandwidth and most realizations are limited to a few $100\,\si{kHz}$ repetition rates  \cite{cerullo_ultrafast_2003,varillas_microjoule-level_2014,manzoni_design_2016,bruder_generation_2021}. Supercontinuum generation in optical fibers provides a viable route for obtaining broadband UV radiation \cite{jiang_deep-ultraviolet_2015,kudlinski_zero-dispersion_2006,stark_extreme_2012,lesko_six-octave_2021} but the multi-octave spectral coverage comes at the cost of low spectral power.

Advances in hollow-core fiber technology have enabled tunable, bright, and broadband radiation at MHz repetition rates. By engineering the waveguide dispersion, resonant dispersive waves (RDW) can efficiently generate deep ultraviolet (DUV) \cite{joly_bright_2011,kottig_generation_2017,suresh_deep-uv-enhanced_2021} and near ultraviolet (NUV) \cite{hosseini_uv_2018,travers_high-energy_2019} radiation. Recently, the favorable temporal structure ($< 3\,\si{fs}$) \cite{ermolov_characterization_2016,reduzzi_direct_2023} and tunability ($236\,\si{nm}$ to $377\,\si{nm}$) of RDW emission were used in time-resolved experiments at kHz repetition rates to perform photoelectron imaging studies \cite{kotsina_spectroscopic_2022} and transient absorption spectroscopy \cite{lee_few-femtosecond_2024}; a beamline for attochemistry studies is under construction \cite{ghezzi_design_2023}.

In this study, the tunability of RDW emission is used to generate UV and visible (VIS) radiation at repetition rates of $5\,\si{MHz}$. The reliability and stability of the ultra-broadband frequency comb source are proven by conducting spectroscopic measurements across the full bandwidth.
$\mathrm{NO_2}$ is used as a spectroscopic prototype due to its broadband, feature-rich absorption response across the UV and VIS regions. It is an atmospheric trace gas of considerable significance, given its role in multiple chemical reactions occurring in the troposphere. Not only is the formation of ozone \cite{muller_impact_2021}, aerosols \cite{liu_effects_2022}, and PM2.5 directly linked to $\mathrm{NO_2}$ concentration, but $\mathrm{NO_2}$ also impacts human health and our surrounding ecosystem. Because one of the two main $\mathrm{NO_2}$ sources is the internal combustion engine of cars, elevated concentrations are particularly observed in urban air during the morning and evening hours, coinciding with traffic rush hours \cite{carslaw_evidence_2005,gupta_dependence_2019}. The second main source of $\mathrm{NO_2}$ is temperature control in buildings, which leads to higher concentrations in winter. Since long-term exposure has been linked to asthma \cite{anenberg_long-term_2022} and high concentrations in a short period also have significant health effects, air quality monitoring is vital. The light source developed here allows to measure trace gases across a broad VIS and NUV spectrum, as demonstrated with $\mathrm{NO_2}$, opening the path toward multi-species detection.

\section{Experimental Methods}
\label{sec:expMeth}

\subsection{Experimental setup}
\label{subsec:expSetup}

\begin{figure}[t!]
    \centering
    \includegraphics[width=1\textwidth]{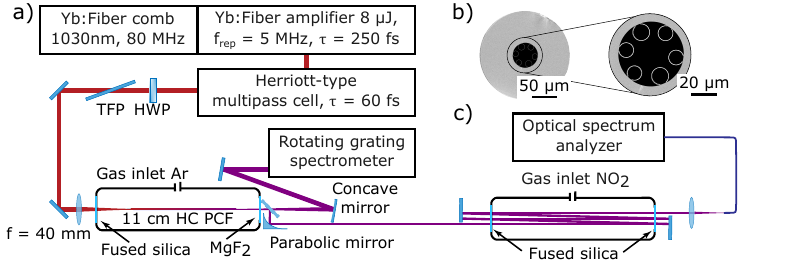}
    \caption{Experimental setup for super-continuum generation and $\mathrm{NO_2}$ absorption spectroscopy. a) The light source is driven by a commercial Yb:fiber frequency comb, seeding a pulse-picking Yb:fiber amplifier. Pulse compression is achieved in a Herriott-type multipass cell. The output power can be adjusted with a half-wave plate (HWP) and a thin film polarizer (TFP). The single-ring hollow-core photonic crystal fiber (HC PCF) is located in a high-pressure gas cell. The output spectrum is monitored using a rotating grating spectrometer. b) Scanning electron microscope cross-section image of the full fiber (left) and the core structure (right). c) The beam traverses a mixture of $\mathrm{N_2}$ and $\mathrm{NO_2}$ in the sample cell with five passes. The transmitted light is detected with an optical spectrum analyzer.}
    \label{fig:expSetup}
\end{figure}

A pulse-picking Yb:fiber frequency comb system with a repetition rate adjustable between $250\,\si{kHz}$ and $80\,\si{MHz}$ delivers $250\,\si{fs}$ pulses with up to $8\,\si{\mu J}$ pulse energy. The pulses are compressed to about $60\,\si{fs}$ by employing a Herriott-type multipass cell with $\mathrm{SiO_2}$ as a nonlinear medium, similar in design to  \cite{fritsch_all-solid-state_2018}. After the multipass cell, beam stabilization is used to minimize pointing variations. The remaining chirp of the broadened pulses is reduced with chirped mirrors (UltraFast Innovations). A half-wave plate and a thin film polarizer attenuate the pulse energy to about $1\,\si{\mu J}$ for optimum frequency conversion. The comb is launched into a gas-filled single-ring hollow-core photonic crystal fiber (SR HC PCF) for generating NUV radiation via RDW emission. The fiber consists of a core with a diameter of $(26.1 \pm 1.9)\,\si{\mu m}$ (slightly elliptical) surrounded by a ring of six capillaries with diameters of $(11.6 \pm 0.9)\,\si{\mu m}$ and a wall thickness of around $250\,\si{nm}$ (\autoref{fig:expSetup}b). In the visible and near-infrared, the evacuated waveguide exhibits anomalous dispersion, which can be tailored by changing the pressure of the filling gas. In this study, the $11\,\si{cm}$ long fiber is filled with $20-40\,\si{bar}$ of argon. The small fiber core diameter allows for the generation of NUV radiation using $1030\,\si{nm}$ pulses carrying energies as low as $(0.9 - 1.2)\,\si{\mu J}$. The light exits the cell through an $\mathrm{MgF_2}$ window followed by a second $\mathrm{MgF_2}$ window at $45^\circ$ used as a beam splitter. The transmitted beam is sent to a rotating grating spectrometer (APE WaveScan) to acquire reference spectra that account for changes in the spectrum due to coupling efficiency fluctuations. The reflected beam passes through the sample cell ($l = 34\,\si{cm}$) five times to increase the interaction path length to $170\,\si{cm}$. The resulting light is coupled into a multimode fiber toward an optical spectrum analyzer (Yokogawa AQ6373E), offering $0.02\,\si{nm}$ spectral resolution from $350$ to $1100\,\si{nm}$.

\subsection{Measurement protocol and referencing}
\label{subsec:expProtocol}

The broad RDW is tuned to six different central wavelengths within the region of interest to determine the absorption across a broad wavelength range. Spectra are acquired in $39\,\si{s}$ each over $10\,\si{nm}$ segments with $0.02\,\si{nm}$ resolution. Each segment is an average of twenty spectra. The measurement sequence is repeated as follows to record the sample and reference spectrum as close in time to each other as possible: 1) measuring a reference spectrum with $10\,\si{nm}$ bandwidth, 2) filling the cell with the sample gas, 3) measuring a $10\,\si{nm}$ broad sample spectrum, 4) measuring a sample spectrum of a second adjacent $10\,\si{nm}$ segment, 5) evacuating the cell, 6) measuring the reference spectrum of the corresponding adjacent $10\,\si{nm}$ segment.
Despite evacuating and refilling the sample cell between measurements, sample consistency is ensured because the gas used is a standard commercial mixture of originally $500\,\si{ppm}$ $\mathrm{NO_2}$ diluted by $\mathrm{N_2}$. The partial pressure in the cell is $(800 \pm 10)\,\si{mbar}$ for all measurements.

A simultaneous reference spectrum is acquired with a rotating grating spectrometer. The final absorbance is calculated by $A = \log_{10}(I_{\mathrm{ref}}/I_\mathrm{samp})$, with $I_{\mathrm{ref}} = I_{\mathrm{OSA,ref}}/I_{\mathrm{RotGrat}}$ and $I_{\mathrm{samp}} = I_{\mathrm{OSA,samp}} /I_{\mathrm{RotGrat}}$, accounting for the fluctuations in the fiber coupling efficiency by normalizing to the reference $I_{\mathrm{RotGrat}}$. A least squares fit of the absorbance computed from literature absorption cross-section data \cite{harder_temperature_1997} to the experimental data determines the $\mathrm{NO_2}$ concentration in the sample cell, with the concentration $c$ as a free variable.

\subsection{Comb stability characterization}
\label{subsec:expComb}
To show that the frequency comb is conserved, a $f-2f$ interferometer is implemented. The experimental setup uses a collinear geometry \cite{baltuska_phase-controlled_2003,kakehata_single-shot_2001}. For this experiment, the amplifier operates at $10\,\si{MHz}$ repetition rate. The argon pressure within the fiber is set to $24\,\si{bar}$ to generate an RDW at a central wavelength of $330\,\si{nm}$. The $f-2f$ interferometry is performed between the frequency-doubled infrared driving field ($1030\,\si{nm}$) and the green tail ($515\,\si{nm}$) of the continuum. When these two combs are superimposed, a beat signal is generated with the frequency of the carrier-envelope-offset frequency ($f_{\mathrm{CEO}}$) \cite{helbing_carrier-envelope_2003}. For this experiment, the collimated light behind the fiber is refocused into an LBO crystal (type I phase matching to the $1030\,\si{nm}$ fundamental beam) to generate its second harmonic. The two cross-polarized combs at $2f$ from fiber and LBO are mixed with a half-wave plate and a polarizing beamsplitter cube. The green part of the spectrum is selected using a transmission grating and a slit and focused onto an amplified Si photodiode (Thorlabs PDA10A2). The beat notes are acquired using an electronic spectrum analyzer without any additional amplification.

\section{Experimental Results}
\label{sec:expRes}

\subsection{Pressure-dependent spectral shifts}
\label{subsec:pressureTuning}

\begin{figure}[t!]
    \centering
    \includegraphics[width=1\textwidth]{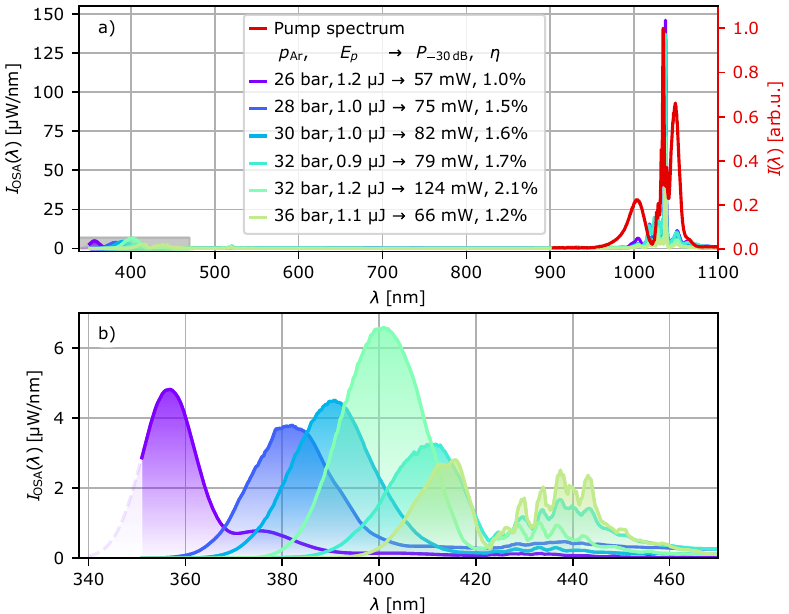}
    \caption{Resonant dispersive wave spectra covering the UV/VIS range. a) Spectrum of the pump with $1030\,\si{nm}$ central wavelength (red). Measured spectra for variable argon pressures $p_{\mathrm{Ar}}$ and launched pulse energies $E_{\mathrm{p}}$ in the fiber. Note the absence of intensity between the resonant dispersive wave and the broadened driving field. b) Closeup of the generated resonant dispersive waves. The power $P_{-30\,\si{dB}}$ provided in the legend is compensated for optical losses at the beamsplitter.}
    \label{fig:spectraOverParams}
\end{figure}

The fiber allows continuous tuning of a $20\,\si{nm}$ (FWHM) broad spectrum from $(300$ to $465)\,\si{nm}$ by changing the argon pressure and launched power. Here the tunability is only shown starting at $350\,\si{nm}$ due to the limited spectral coverage of the available optical spectrum analyzer, covering $350\,\si{nm}$ to $1100\,\si{nm}$. As the argon pressure increases, the central wavelength of the RDW shifts to longer wavelengths due to phase-matching \cite{travers_ultrafast_2011}. Combining pressure changes with adjustments to the launched pulse energy, the NUV radiation can be spectrally tuned as shown in \autoref{fig:spectraOverParams}. The average power in the UV is about $80\,\si{mW}$ ($-30\,\si{dBc}$ width), resulting in an average conversion efficiency of $1.5\,\%$ (\autoref{fig:spectraOverParams}a). The pulse energy of the UV is about $16\,\si{nJ}$. As shown in \cite{ermolov_characterization_2016,reduzzi_direct_2023,brahms_direct_2019}, due to the underlying fiber dynamics, the dispersive wave emission can additionally be ultrashort in time ($<3\,\si{fs}$), enabling experiments with high temporal resolution \cite{kotsina_spectroscopic_2022,lee_few-femtosecond_2024}. While optical parametric amplifier based systems can also generate broad spectra in the UV at kHz repetition rates resulting in $\approx 8\,\si{fs}$ pulses \cite{varillas_microjoule-level_2014,bruder_generation_2021,baum_tunable_2004}, this RDW method allows scaling above MHz repetition rates required for dual-comb spectroscopy or any other high signal-to-noise ratio applications \cite{cerullo_ultrafast_2003,manzoni_design_2016}.

\subsection{Broadband absorption spectroscopy of \texorpdfstring{$\mathrm{NO_2}$}{NO2}}
\label{subsec:NO2absSpec}

\begin{figure}[t!h]
    \centering
    \includegraphics[width=1\textwidth]{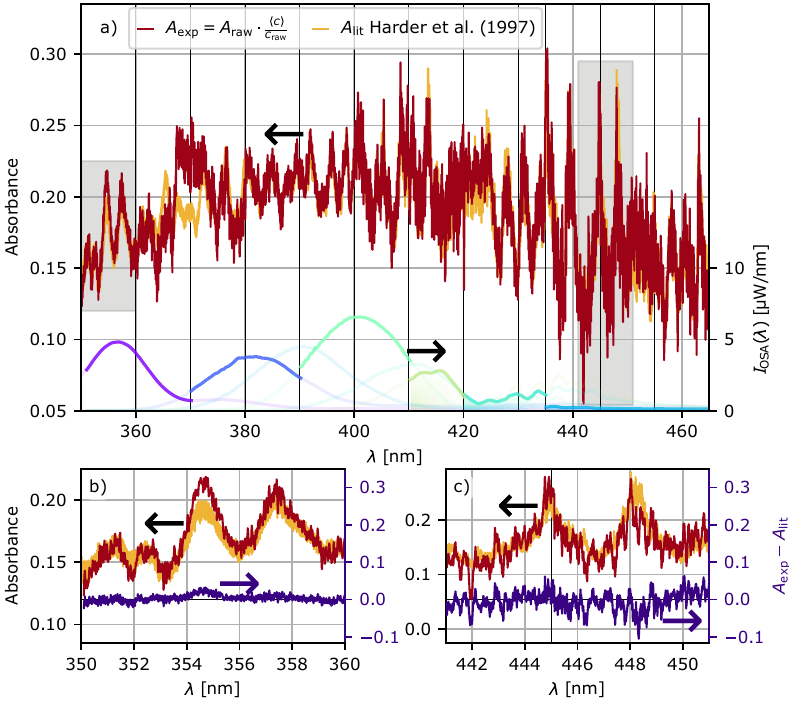}
    \caption{Absorption spectrum of $\mathrm{NO_2}$. a) Measured absorbance (red) of $\mathrm{NO_2}$ in the sample cell. The acquired spectra are multiplied with a factor to correct concentration changes. The x-axis was converted to vacuum wavelength using the updated Edlén formula \cite{birch_updated_1993}. The result is in good agreement with the literature  \cite{harder_temperature_1997} (orange). The spectra used for the acquisition are shown at the bottom. b) Close-up of absorbance obtained with the resonant dispersive wave between $350\,\si{nm}$ and $360\,\si{nm}$. The standard deviation of the residual shown in purple is $0.0087$. c) Close-up of absorbance performed with the tail of the spectrum between $441\,\si{nm}$ and $451\,\si{nm}$. The standard deviation of the residual is $0.0248$.}
    \label{fig:no2Absorption}
\end{figure}

The spectroscopic capabilities of the tunable and broadband light source are demonstrated by investigating the absorption of $\mathrm{NO_2}$. This gas features a large set of small absorption features across the NUV on top of a pronounced trend.
The measured absorbance of $\mathrm{NO_2}$ is shown in red in \autoref{fig:no2Absorption}. The measured spectra were multiplied by a varying factor to correct the absorbance to the mean observed sample concentration of $(235 \pm 59)\,\si{ppm}$ of $\mathrm{NO_2}$. The factors and raw data are provided in Supplement 1. The updated Edlén formula \cite{birch_updated_1993} was used to convert the x-axis to vacuum wavelengths. In \autoref{fig:no2Absorption}, a high-resolution absorption cross-section literature curve is plotted in orange  \cite{harder_temperature_1997} for comparison. The literature data was acquired using an Xe arc lamp and a Fourier transform spectrometer.

The presented measurement matches almost all absorption features of $\mathrm{NO_2}$. The deviation at $370\,\si{nm}$ can be explained by an observed drift of the laser pointing, which changed the spectrometer coupling and led to a baseline change. In \autoref{fig:no2Absorption}b, a close look at the feature aligned with the $357\,\si{nm}$ RDW spectrum displays the high-resolution capabilities of the technique. In purple, the residual shows a standard deviation of $0.0087$. In \autoref{fig:no2Absorption}c, the absorbance is acquired with the tail of the RDW spectrum. In this case, the RDW has its maximum at $390\,\si{nm}$, far from the $\mathrm{NO_2}$ features. Features like the double peak at $448\,\si{nm}$ are still resolved, even with the spectral intensity being more than an order of magnitude lower. The standard deviation of the residual is approximately three times higher than the value observed in \autoref{fig:no2Absorption}b. The noise can be attributed to the emergence of spurious features, such as the modulation observed around $442\,\si{nm}$. These are the result of a minor shift in the position of a peak in the fiber spectrum due to a change in HC PCF incoupling. The generated spectrum is less smooth in this region due to the higher soliton order of the driving pulses and the spectral anti-crossing between the guided mode and the core-wall resonances.

\subsection{Comb structure of generated spectrum}
\label{subsec:combStruct}

\begin{figure}
    \centering
    \includegraphics[width=1\textwidth]{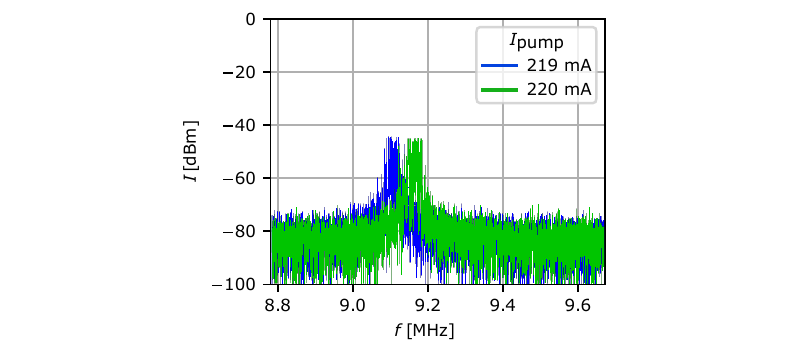}
    \caption{Preservation of comb structure by the fiber. Beat note created by interference of the second harmonic and the tail of the supercontinuum for two cavity dispersion settings, set by different oscillator pump currents. It is about $100\,\si{kHz}$ broad ($-20\,\si{dB}$), measured with a resolution bandwidth of $10\,\si{kHz}$.}
    \label{fig:f-2fBeat}
\end{figure}

The RDW is driven with a frequency comb. To confirm that the comb structure is conserved during the up-conversion, $f-2f$ interferometry is used. The resulting beat signal is shown in \autoref{fig:f-2fBeat}. To verify that the observed beat is caused by $f_{\mathrm{CEO}}$, the cavity dispersion is changed by setting a different oscillator pump current, leading to the expected shift of the beat frequency ($219\,\si{mA}$ (blue) vs. $220\,\si{mA}$ (green)). The seed oscillator is specified to be below $200\,\si{kHz}$ $f_{\mathrm{CEO}}$ beat linewidth. Based on the observed $f_{\mathrm{CEO}}$ beatnote width of $100\,\si{kHz}$ at $-20\,\si{dB}$, an upper boundary for the increase of linewidth of $100\,\si{kHz}$ can be concluded. It should be noted that this figure represents the complete chain from oscillator to pulse picking, amplification, pulse compression, and broadening. These stages are not fully optimized yet for phase-locked operation and are known to introduce various amounts of phase noise \cite{baltuska_phase-controlled_2003,vries_acousto-optic_2015,mei_space-time_2021,schuster_ultraviolet_2021}.

The stability of the source was ensured by selecting the experimental parameters and the species of the gas filling the fiber to minimize its ionization and thus avoid long-lived effects, which can impair the pulse-to-pulse stability \cite{kottig_generation_2017,koehler_post-recombination_2021}. Numerical simulations of the nonlinear propagation of the pulses in the HC-PCF based on the code described in \cite{tani_multimode_2014} and using the Perelomov, Popov, Terent’ev ionization rate (as described in \cite{couairon_femtosecond_2007}) reveal that for the selected parameters the peak-free electron density remains below $10^{12} \,\si{electrons/{cm}^3}$, which is sufficiently low to avoid significant ionization-induced thermal effects \cite{koehler_post-recombination_2021}.
In a similar study involving RDW frequency conversion with an $f_{\mathrm{CEO}}$ stabilized laser system, the $f_{\mathrm{CEO}}$ variations below $240\,\si{Hz}$ were characterized and it was concluded that the phase variation is about $330\,\si{mrad}$ RMS over $2.5$ minutes \cite{ermolov_carrier-envelope-phase-stable_2019}. The results of the present study indicate that the comb stability is also not impacted by high-frequency noise.

\FloatBarrier

\section{Conclusion}
\label{sec:conclusion}

In this work, a broadband tunable laser frequency comb source is demonstrated with a repetition rate of $5\,\si{MHz}$, covering more than $115\,\si{nm}$ ($240\,\si{THz}$) across the UV/VIS spectrum. We utilized the broadband source to measure the detailed absorption of $\mathrm{NO_2}$ from $350\,\si{nm}$ to $465\,\si{nm}$. The sample concentration was determined to be $(235 \pm 59)\,\si{ppm}$.
Additionally, our measurements show that the comb structure is conserved, even at a repetition rate of $10\,\si{MHz}$.
These results make ultra-broadband frequency comb spectroscopy in the UV/VIS attainable. A second laser system to perform comparable broadband measurements with the additional benefits of dual-comb spectroscopy is currently under construction. The broad spectral coverage will enable multi-species trace gas sensing throughout the NUV. Another application could be the tracking of reaction pathways. With the high photon flux, free space trace gas measurement campaigns over km path lengths, usually challenging due to high optical losses, come within reach.

\textbf{Funding.} HORIZON EUROPE European Research Council (947288); Austrian Science Fund (Y1254), EIC Pathfinder 101046424 - TwistedNano.\\
\textbf{Disclosures.} The authors declare no conflicts of interest.\\
\textbf{Data availability.} Data underlying the results presented in this paper are not publicly available at this time but may be obtained from the authors upon reasonable request.\\
\textbf{Supplemental document.} See Supplement 1 for supporting content.

\makeatletter
\renewcommand \thesection{S\@arabic\c@section}
\renewcommand\thetable{S\@arabic\c@table}
\renewcommand \thefigure{S\@arabic\c@figure}
\makeatother

\setcounter{figure}{0}
\setcounter{section}{0}

\section{Supplementary materials: Ultra-broad\-band UV/ VIS spectroscopy enabled by resonant dispersive wave emission of a frequency comb}
\label{sec:supplement}

\subsection{Correction of concentration variations}
\label{subsec:corrSup}

\begin{figure}[th]
    \centering
    \includegraphics[width=1\textwidth]{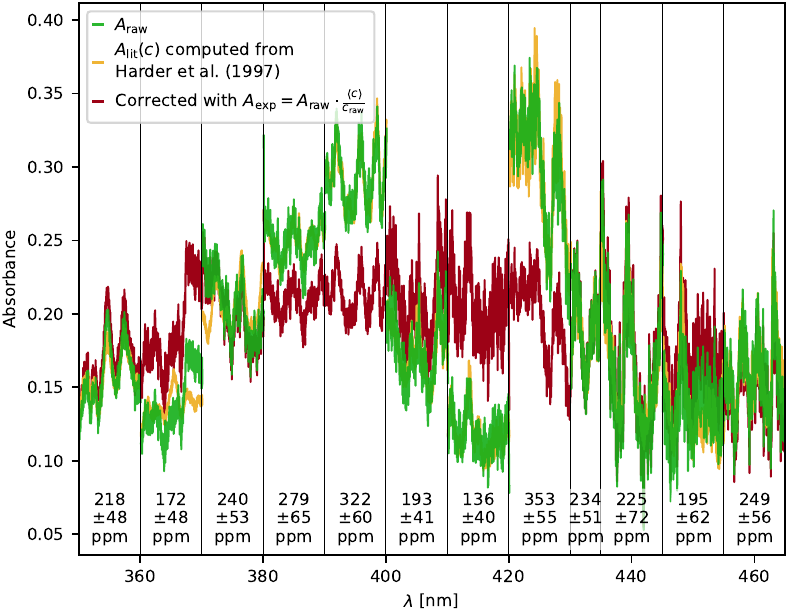}
    \caption{Raw absorbance data and correction steps applied to arrive at \autoref{fig:no2Absorption}. The raw measured data is shown in green. From it, the concentration of the sample is extracted using a least squares fit shown in orange, using literature data from \cite{harder_temperature_1997}. In red, the corrected absorbance is shown, as it is plotted in \autoref{fig:no2Absorption}. The average sample concentration is $(235 \pm 59)\,\si{ppm}$. See \autoref{tab:corrFactors} for the applied correction factors.}
    \label{fig:corrSteps}
\end{figure}

The measurement is performed by scanning multiple $10\,\si{nm}$ consecutively (\autoref{subsec:expProtocol}). Due to the effects of measurement time, the sample concentration decreases. The averages of the raw measured absorbance traces are shown in \autoref{fig:corrSteps} (green). To extract the sample concentration, a least squares fit to the raw measured absorbance traces is performed with the concentration $c$ as the free variable. The resulting average value is shown at the bottom of the figure and in \autoref{tab:corrFactors}. In orange, the literature absorbance corresponding to the extracted average concentration for each interval is shown, demonstrating excellent agreement. The outliers were already discussed in \autoref{subsec:NO2absSpec}. To arrive at the absorbance for the average sample concentration shown in red, the raw absorbance is multiplied by $\langle c \rangle / c_{\mathrm{raw}}$, where $\langle c \rangle$ is the average sample concentration of all scans and $c_{\mathrm{raw}}$ is the sample concentration observed during the measurement of a $10\,\si{nm}$ interval. The values of the fraction are listed in \autoref{tab:corrFactors}.

\begin{table}[bh]
\caption{Extracted sample concentrations and correction factors.}
\begin{tabular}{p{.7cm}p{.48cm}p{.48cm}p{.48cm}p{.48cm}p{.48cm}p{.48cm}p{.48cm}p{.48cm}p{.48cm}p{.48cm}p{.48cm}p{.48cm}}
\toprule
Cen\-tral wavelength [nm] & 355      & 365      & 375      & 385      & 395     & 405      & 415      & 425      & 435      & 440      & 450      & 460      \\
\midrule
$c$ [ppm]                   & $218 \pm 48$ & $172 \pm 48$ & $240 \pm 53$ & $279 \pm 65$ & $322 \pm 60$ & $193 \pm 41$ & $136 \pm 40$ & $353 \pm 55$ & $234 \pm 51$ & $225 \pm 72$ & $195 \pm 62$ & $249 \pm 56$ \\
Factor                        & 1.08     & 1.36     & 0.98     & 0.84     & 0.73    & 1.21     & 1.72     & 0.66     & 1.00     & 1.04     & 1.21     & 0.94     \\
\bottomrule
\label{tab:corrFactors}
\end{tabular}
\end{table}
\FloatBarrier
\printbibliography
\end{document}